\documentclass[12pt,a4paper]{ijicic}

\usepackage{amsmath}
\usepackage{amssymb}
\usepackage[dvips]{graphicx}
\usepackage{epsfig}
\usepackage{epsf}
\include{figure}
\usepackage{verbatim}
\usepackage{paralist}
\usepackage{multirow}

\title{}
\title[FUZZY FEEDBACK SCHEDULING OF EMBEDDED CONTROL SYSTEMS]
      {FUZZY FEEDBACK SCHEDULING OF RESOURCE-CONSTRAINED EMBEDDED CONTROL SYSTEMS}

\author[F. Xia, Y. Sun, Y.-C. Tian, M. O. Tade and J. Dong]{}

\begin{document}

\maketitle

\begin{center}
\normalsize{\scshape Feng Xia$^{1,2}$, Youxian Sun$^3$, Yu-Chu
Tian$^{2,*}$, Moses O. Tade$^4$ \\* and Jinxiang Dong$^1$}
\footnotesize\rmfamily
\medskip

$^1$College of Computer Science and Technology\\
Zhejiang University\\
Hangzhou 310027, P. R. China\\
f.xia@ieee.org
\medskip

$^2$Faculty of Information Technology \\
Queensland University of Technology\\
GPO Box 2434, Brisbane QLD 4001, Australia\\
y.tian@qut.edu.au\\
$^*$Corresponding author
\medskip

$^3$State Key Laboratory of Industrial Control Technology\\
Zhejiang University\\
Hangzhou 310027, P. R. China
\medskip

$^4$Department of Chemical Engineering\\
Curtin University of Technology\\
GPO Box U1987, Perth WA 6845, Australia

\end{center}

\medskip
\centerline{\textsf{Received August 2007; revised January 2008}}
\medskip

\begin{abstract}
{\em The quality of control (QoC) of a resource-constrained embedded
control system may be jeopardized in dynamic environments with
variable workload. This gives rise to the increasing demand of
co-design of control and scheduling. To deal with uncertainties in
resource availability, a fuzzy feedback scheduling (FFS) scheme is
proposed in this paper. Within the framework of feedback scheduling,
the sampling periods of control loops are dynamically adjusted using
the fuzzy control technique. The feedback scheduler provides QoC
guarantees in dynamic environments through maintaining the CPU
utilization at a desired level. The framework and design methodology
of the proposed FFS scheme are described in detail. A simplified
mobile robot target tracking system is investigated as a case study
to demonstrate the effectiveness of the proposed FFS scheme. The
scheme is independent of task execution times, robust to measurement
noises, and easy to implement, while incurring only a small
overhead.}\\
{\bf Keywords:} Feedback scheduling, Fuzzy logic control, Embedded
control systems, Resource management, Mobile robot
\end{abstract}

\section{Introduction}
Despite their popularity, embedded systems are typically resource
limited \cite{R1,R2,R3}. For instance, there are usually constraints on the
processing speed, memory size, and communication bandwidth. As the complexity of various applications grows
continuously, multiple tasks have to compete for the limited
processor resource in many cases. In this context, the overall
quality-of-control (QoC) of an embedded control system depends not
only on the design of control algorithms, but also on the scheduling
of shared computing resources. With traditional open-loop scheduling
schemes, however, the temporal attributes of these systems will be
significantly affected by workload variations. This may potentially
cause the overall QoC to deteriorate \cite{R4,R5}.

Feedback scheduling offers a promising approach to flexible QoC
management in dynamic environments with variable workload \cite{R1,R3}. The
basic idea of feedback scheduling is to allocate available resources
online among tasks by adapting their timing parameters, e.g.,
periods, based on feedback information about the actual resource
utilization. By exploring the integration of feedback control and
real-time scheduling, feedback scheduling can deal with
uncertainties in resource availability, and thus provide QoC
guarantees for the system. In the last decade, much effort has been
made in this area, for example, \cite{R5,R6,R7,R8,R9,R10,R11,R12,R13,R14,R15}.
Conventional feedback scheduling schemes, particularly the most
widely studied optimal feedback scheduling \cite{R6,R7,R8}, require that the
following information be available at runtime: 1) the control cost
function as a function of sampling period for each control loop; 2)
the execution times of control tasks; and 3) the CPU utilization. In
practical systems, however, some of this information may not always
be available. In many complex control systems, it is difficult, if
not impossible, to describe the performance index of QoC as a
function of sampling period. In a variety of
commercial off-the-shelf (COTS) operating systems, direct
measurement of task execution times is not always supported \cite{R12}.
Measurement noises inevitably exist in real-world applications, even
when online measurement of task execution times is supported by the
underlying operating system kernel. Available measurements or
estimations of the system workload are consequently imprecise. When
the availability and accuracy of critical information cannot be
guaranteed, conventional feedback scheduling schemes may not perform
as expected or even become ineffective in some circumstances.

This paper considers resource-constrained embedded control systems
in which 1) the runtime availability of some important information
such as task execution times cannot be guaranteed; and 2) the system
parameter measurements are imprecise. To provide QoC guarantees in
dynamic environments, this paper will propose a fuzzy feedback
scheduling (FFS) scheme that takes advantage of both fuzzy logic
control \cite{R16,R17,R18,R19} and feedback scheduling. A mapping from
feedback scheduling to fuzzy logic control will be built. Through
dynamically adjusting the periods of control tasks, the fuzzy
feedback scheduler attempts to maintain the CPU utilization at a
desired level, thus enabling flexible QoC management.

The use of the fuzzy control technique in feedback scheduling is
mainly inspired by its powerful capability in dealing with
nonlinearity, imprecision, and uncertainty. Since direct mathematical modelling of the
controlled process is not required for fuzzy control system design,
the proposed scheme does not depend on explicit formulation of the
relationship between control cost functions and sampling periods. As
a formal methodology to emulate the intelligent decision-making
process of a human expert, fuzzy control provides a simple and
flexible way to reach a definite conclusion based on imprecise,
noisy, or incomplete input information. Consequently, the FFS scheme
is independent of task execution times, and capable of handling
uncertainties, imprecision, and incompletion of system parameters.
Since fuzzy controllers have a simple structure, the proposed scheme
is easy to implement with only a small computational overhead.

Applying fuzzy control techniques to resource management in
general-purpose computing and communication systems has attracted
increasing attention, e.g., \cite{R10,R20}. However, none of these papers
have dealt with control applications where the QoC is the main
concern. Jin and colleagues \cite{R11} employed fuzzy logic in scheduling
control tasks. However, feedback scheduling based on fuzzy logic
control remains unexplored for real-time control tasks. In our
previous work \cite{R15}, a fuzzy feedback scheduler has been developed to
manipulate the execution times of anytime control algorithms; however, it
does not deal with anytime control algorithms. The present work substantially extends our
preliminary work reported in \cite{R13,R14}. Instead of task
execution times, the sampling periods will be chosen as the manipulated
variables in this work. A generalized framework of
fuzzy feedback scheduling and detailed design procedures for the
fuzzy feedback scheduler will be described along with
a case study. To reduce the computational overhead, this paper
employs the look-up table method, which has not been explored in
\cite{R13,R14}.

The paper is organised as follows. The problem to be addressed is
described in Section \ref{sec:2}. Section \ref{sec:3} presents the basic framework of
fuzzy feedback scheduling. In Section \ref{sec:4}, the design procedures for
fuzzy feedback scheduler are discussed in detail. Section \ref{sec:5}
evaluates the performance of fuzzy feedback scheduling and compares
the results with those of the traditional open-loop scheduling
scheme and an ideal feedback scheduling scheme. The paper is
concluded in Section \ref{sec:6}.

\section{Problem Statement}
\label{sec:2}

Consider a system in which \emph{N} independent control tasks are
running concurrently on a processor with limited processing power.
Each control task executes a well-designed control algorithm, which
is responsible for the control of a physical process. In addition to
these control tasks, some non-control tasks for, e.g. data backup,
human machine interaction, and remote communication, may also
exist. For simplicity, all tasks in the system are assumed to be
periodic. The timing parameters of control task \emph{i} are defined
below:
\begin{itemize}
\item $h_i$: task period, which is equal to the sampling period of the
corresponding control loop, and is available precisely online.
\item $c_i$: task execution time, which is time-varying and unavailable at runtime.
\end{itemize}

Without loss of generality, assume that the measurement of CPU
utilization is calculated by:
\begin{equation}
\label{equ:1}
\hat{U}=\sum^N_{i=1}\frac{c_i}{h_i}+U_{others}+\delta_U
\end{equation}
where $U_{others}$ represents the total CPU utilization of all
non-control tasks, and $\delta_U$ is measurement noise. It is worth
mentioning that the design of a fuzzy feedback scheduler does not
rely on how to compute the CPU utilization measurement. In the
system considered here, $c_i$ and $U_{others}$ are unavailable to
the feedback scheduler. The information available to the feedback
scheduler includes the utilization measurement $\hat{U}$ and the
sampling periods $h_i$.

Since the task execution times vary over time, it is possible that
the system becomes overloaded. In this situation, the overall QoC
deteriorates and the system may even become unstable [5]. On the
other hand, when the system is underloaded, some computing resources will be
wasted, thus yielding worse performance than possible. Therefore,
flexible QoC management should be provided in this dynamic
environment so as to guarantee the QoC in overloaded conditions and
to improve the overall QoC in the presence of light workload through
making better use of available resources. Accordingly, the problem
to be addressed can be stated as follows: \emph{Given a set of
control tasks with time-varying and unknown execution times and a
processor with limited computing resources, design a feedback
scheduler to dynamically allocate available resources among these
control tasks such that flexible QoC management is achieved in the
presence of measurement noises.} Due to the unavailability of task
execution times, conventional feedback scheduling strategies that
depend on complete formulation of task models become inapplicable.

\section{Fuzzy Feedback Scheduling Framework}
\label{sec:3}

To address the
above problem, this section proposes an intelligent feedback
scheduling scheme based on the fuzzy control technique. Figure 1
depicts the architecture of the proposed FFS scheme. In addition to
the control loops, an outer feedback loop is introduced to implement
feedback scheduling. The basic role of the feedback scheduler is to
adjust the periods of the control tasks dynamically to achieve a
desired level of CPU utilization. The timing attributes of all
non-control tasks cannot be changed by the feedback scheduler.

\begin{figure}[htbp!]
\centering
\includegraphics[scale=0.7, bb=161 359 520 517]{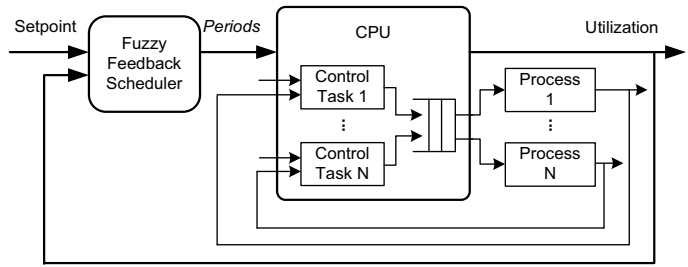}
\caption{Architecture of fuzzy feedback scheduling.}
\label{fig1}
\end{figure}

From the control perspective, the fuzzy feedback scheduler is a
fuzzy controller with CPU utilization being the controlled variable
and sampling periods being the manipulated variables. It is
intuitive that the desired CPU utilization $U_R$ should not violate
the schedulability constraint associated with the system. In
practice, some schedulability margin must be preserved when choosing
$U_R$ because of the presence of measurement noises.

To simplify the design of the feedback scheduler, a simple
rescaling method is employed to adjust the sampling periods:
\begin{equation}
\label{equ:2}
h_i(j)=\eta(j)h_i(j-1),\ \ \ i=1,\cdots,N
\end{equation}
where $\eta(j)$  is the \emph{period rescaling factor}, and \emph{j}
denotes the invocation instant of feedback scheduler. In \cite{R6} Cervin
\emph{et al} have used a similar method, which delivers good scheduling performance.

Taking into account the maximum allowable sampling period
$h_{i,max}$, the calculation of the sampling periods is expressed as:
\begin{equation} \label{equ:3}
\begin{split}
h_i(j)=\mbox{min}\{h_{i,max},\ \eta(j)h_i(j-1)\},\ \ \ i=1,\cdots,N
\end{split}
\end{equation}

Ideally, if the timing attributes of all tasks are precisely known,
the period rescaling factor can be set to:
\begin{equation}
\label{equ:4}
\eta(j)=\frac{\sum^N_{i=1}\frac{c_i(j)}{h_j(j-1)}}{U_R-U_{others}(j)}
\end{equation}

After the sampling periods are altered using (\ref{equ:3}) and
(\ref{equ:4}), the total CPU utilization of the control tasks in the
next invocation interval becomes:
\begin{equation} \label{eq:5}
\begin{split}
U(j)&=\sum^N_{i=1}\frac{c_i(j)}{h_i(j)}+U_{others}(j)
=\frac{1}{\eta(j)}\sum^N_{i=1}\frac{c_i(j)}{h_i(j-1)}+U_{others}(j)\\
&=\frac{U_R-U_{others}(j)}{\sum^N_{i=1}\frac{c_i(j)}{h_j(j-1)}}\sum^N_{i=1}\frac{c_i(j)}{h_i(j-1)}
+U_{others}(j)=U_R
\end{split}
\end{equation}

Using this \emph{ideal} method, one can easily achieve the desired
level of CPU utilization in the next invocation interval. However,
this is only true for the ideal case where all necessary system
parameters are precisely known. In the systems described in Section
2, it is impossible to use (\ref{equ:4}) to compute the period
rescaling factor. In this paper the fuzzy control technique is
employed to determine the period rescaling factor $\eta$.

The work flow of the fuzzy feedback scheduling system can be
outlined as follows. At every invocation instant, the feedback
scheduler samples the CPU utilization that the system has been
monitoring from the previous invocation instant, and compares it
with its desired value. Based on the control error of the CPU
utilization and the change in the control error, the fuzzy controller acting as
the feedback scheduler produces a corresponding period rescaling
factor. The sampling period of each control loop is then re-assigned
using (\ref{equ:3}).

\section{Fuzzy Feedback Scheduler Design}
\label{sec:4}

This section will describe in detail the
design procedures of the fuzzy feedback scheduler. Throughout the
description, a simplified mobile robot target tracking system is
used as the target application.

\subsection{A case study}Consider a simplified mobile robot system \cite{R21}
as shown in Figure 2. The robot is treated as a point $(x, y)$ on
the plane. It can move on $x$-axis and $y$-axis freely and
independently. The coordinate of the robot on each axis, i.e., $x$
or $y$, is respectively controlled by a separate control loop. The
overall goal of the system control is to track as closely as
possible a mobile target, which is also modelled as a mobile point
on the plane.

The transfer functions of both control loops are $G(s) =
1000/(0.5s^2+s)$. Controllers are designed by discretizing
continuous-time controllers that use the PID
(proportional-integral-derivative) algorithm. The controller
settings are taken from \cite{R4}. Two control tasks, denoted by $\tau_1$
and $\tau_2$, respectively, together with a third periodic and
non-control task $\tau_3$, are running on one processor
concurrently.

\begin{figure}[htb!]
\begin{center}
\includegraphics[scale=0.8, bb=96 418 406 558]{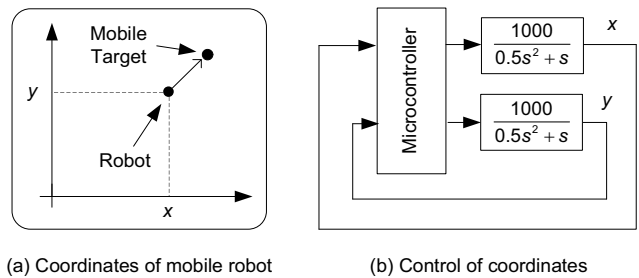}
\caption{A simplified mobile robot system.}
\label{fig2}
\end{center}
\end{figure}

\subsection{Design methodology}

Figure 3 shows the internal structure of the fuzzy
feedback scheduler, where the component of the sampling period
adjustment is omitted for simplicity. Like almost all fuzzy
controllers, the fuzzy feedback scheduler consists of four main
components, i.e., a fuzzification interface, a rule-base, an
inference mechanism, and a defuzzification interface. Inputs to the
fuzzy feedback scheduler are the control error of the CPU utilization,
$e(j) = U_R-\hat{U}(j)$, and the change in the control error, $ec(j) =
e(j)-e(j-1)$; while the output is naturally the sampling period
rescaling factor $\eta$. The inner work flow of the fuzzy feedback
scheduler is as follows. Firstly, the fuzzification interface
translates numeric inputs $e(j)$ and $ec(j)$ into fuzzy sets
characterizing linguistic variables $E$ and $EC$, respectively. The
inference mechanism then activates a predetermined set of linguistic
rules in the rule-base with respect to these linguistic variables,
and produces the fuzzy sets of the output linguistic variable $RF$.
Finally, the defuzzification interface converts the fuzzy
conclusions, which are reached by the inference mechanism, to a
numeric value $\eta(j)$.
\begin{figure}[htb!]
\begin{center}
\includegraphics[scale=0.8, bb=113 339 394 548]{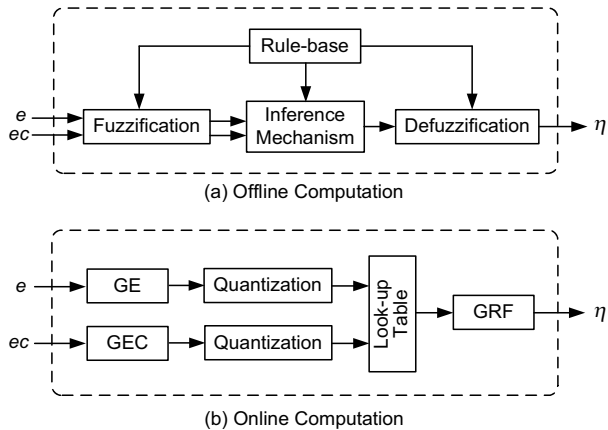}
\caption{Internal structure of fuzzy feedback scheduler.}
\label{fig3}
\end{center}
\end{figure}

To simplify online computations, the look-up table method is
employed to implement the fuzzy feedback scheduler. As shown in
Figure 3(b), the procedures of using this method are as follows.
Firstly, construct a fuzzy control table, which is also called \emph{look-up
table}, via offline computations using the original fuzzy feedback
scheduler depicted in Figure 3(a), and store the table in the
memory. During runtime, the resulting feedback scheduler quantizes
the inputs $e(j)$ and $ec(j)$ using the input scaling factors $GE$
and $GEC$, and then searches in the fuzzy control table to find out
the quantized output value corresponding to the quantized inputs.
The final output $\eta(j)$ is generated by multiplying the quantized
output value with the output scaling factor $GRF$. A key step in
this method is to build a good fuzzy control table. The design
procedures of the fuzzy feedback scheduler for the simplified mobile
robot system are detailed below.

\begin{asparaenum}
\item \emph{Specify the structure of the fuzzy feedback scheduler}

From the above descriptions, a two-dimension fuzzy controller is
used, which has two input variables (\emph{e} and \emph{ec}) and one
output $\eta$. The universes of discourse for $e$, $ec$, and $\eta$
are chosen to be [-0.3, 0.3], [-0.3, 0.3], and [0.5, 1.5],
respectively.

\item \emph{Describe inputs and outputs linguistically}

The sets of linguistic values for the linguistic variables $E$ and
$EC$ are {NB, NS, ZE, PS, PB}, and the set of linguistic values for
$RF$ is {NB, NM, NS, ZE, PS, PM, PB}, where NB represents
\emph{negative big}, NM represents \emph{negative medium}, NS
represents \emph{negative small}, ZE represents \emph{zero}, PS
represents \emph{positive small}, PM represents \emph{positive
medium}, and PB represents \emph{positive big}. To facilitate the
construction of the look-up table, both inputs and output should be
quantized. The sets of quantized values for the two inputs are \{-6,
-5, -4, -3, -2, -1, 0, 1, 2, 3, 4, 5, 6\}. The set of quantized
values for the output is \{-7, -6, -5, -4, -3, -2, -1, 0, 1, 2, 3,
4, 5, 6, 7\}. Accordingly, $GE = 6/0.3 = 20$, $GEC = 6/0.3 = 20$,
and $GRF = 1/14 = 0.0714$.
\item \emph{Specify membership functions for linguistic values}

Figure \ref{fig4} depicts the membership functions used in this
paper for all linguistic values of both input and output linguistic
variables.
\begin{figure}[htb!]
\begin{center}
\includegraphics[scale=0.65, bb=101 288 476 566]{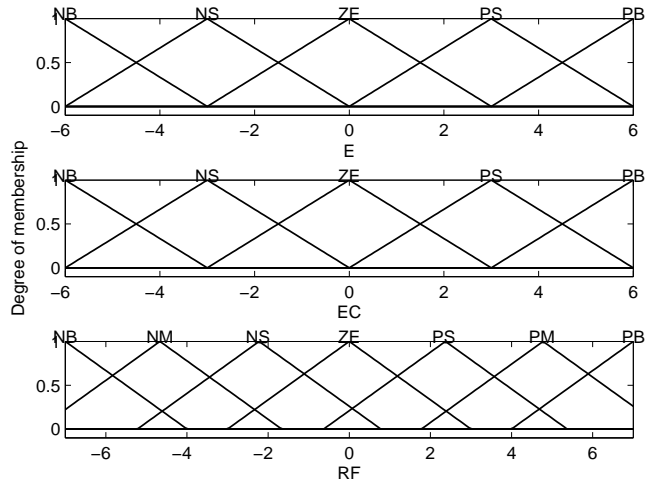}
\caption{Input and output membership functions.}
\label{fig4}
\end{center}
\end{figure}

\item \emph{Determine fuzzy control rules and inference mechanism}

The linguistic rules are established through analyzing the system
behaviour under various conditions. The fuzzy control rules are
constructed as follows. If the measured CPU utilization exceeds the
desired level significantly, a big sampling period rescaling factor
should be used so that the sampling periods are enlarged quickly enough
to avoid too many deadline misses. If the measured CPU utilization
is lower than the desired level, sampling periods should be
decreased slowly to reduce the possibility of overloading. As shown
in Table 1, a total of 25 linguistic rules are built in this paper
for the mobile robot system.

For the inference mechanism, the max-min method is adopted. In the
defuzzification interface, the most popular \emph{centre} of
\emph{gravity} method is used to produce a real number in the
universe of discourse of the output.
\begin{table}[htb!]
\centering
\tabcolsep 3mm
\caption{Linguistic rules.}
\label{table:1}
\begin{center}
\begin{tabular}[b]{|l|l|c|c|c|c|c|}
\hline
\multicolumn{2}{|c|}{}
&\multicolumn{5}{c|}{\parbox{8mm}{$EC$}} \\\cline{3-7}
\multicolumn{2}{|c|}{\raisebox{1.5ex}[0cm][0cm]{\parbox{5mm}{$RF$}}} & \parbox{5mm}{NB} & \parbox{5mm}{NS} & \parbox{5mm}{ZE} & \parbox{5mm}{PS} & \parbox{5mm}{PB} \\\hline
\multirow{5}{3mm}{$E$}
& \parbox{5mm}{NB} & \parbox{5mm}{PB} & \parbox{5mm}{PB} & \parbox{5mm}{PB} & \parbox{5mm}{PB} & \parbox{5mm}{PM}  \\ \cline{2-7}
& \parbox{5mm}{NS} & \parbox{5mm}{PB} & \parbox{5mm}{PB} & \parbox{5mm}{PM} & \parbox{5mm}{PS} & \parbox{5mm}{ZE}  \\ \cline{2-7}
& \parbox{5mm}{ZE} & \parbox{5mm}{PM} & \parbox{5mm}{PS} & \parbox{5mm}{ZE} & \parbox{5mm}{ZE} & \parbox{5mm}{NS} \\ \cline{2-7}
& \parbox{5mm}{PS} & \parbox{5mm}{PS} & \parbox{5mm}{ZE} & \parbox{5mm}{ZE} & \parbox{5mm}{NS} & \parbox{5mm}{NM} \\ \cline{2-7}
& \parbox{5mm}{PB} & \parbox{5mm}{ZE} & \parbox{5mm}{NS} & \parbox{5mm}{NM} & \parbox{5mm}{NB} & \parbox{5mm}{NB} \\\hline
\end{tabular}
\end{center}
\end{table}

\item \emph{Create look-up table}

The final step for offline design of fuzzy feedback scheduler is to generate the
fuzzy control table that can be used online, as given in Table 2.
For this purpose, the quantized value of the output corresponding to
each pair of quantized values of the input variables is calculated. Once
the look-up table is constructed, it can be used at runtime within the framework
given in Figure 3(b).

\begin{table}[htb!]
\centering
\tabcolsep 2mm
\caption{Look-up table for fuzzy feedback scheduler.}
\label{table:2}
\begin{center}
\begin{tabular}[b]{|r|r|r|r|r|r|r|r|r|r|r|r|r|r|r|}
\hline
\multicolumn{2}{|c|}{}
&\multicolumn{13}{c|}{\parbox{8mm}{$ec$}} \\\cline{3-15}
\multicolumn{2}{|c|}{\raisebox{1.5ex}[0cm][0cm]{\parbox{4mm}{$\eta$}}} & \parbox{4mm}{-6} & \parbox{4mm}{-5} & \parbox{4mm}{-4} & \parbox{4mm}{-3} & \parbox{4mm}{-2} & \parbox{4mm}{-1}& \parbox{4mm}{0}& \parbox{4mm}{1}& \parbox{4mm}{2}& \parbox{4mm}{3}& \parbox{4mm}{4}& \parbox{4mm}{5}& \parbox{4mm}{6} \\\hline
\multirow{13}{3mm}{$e$}
& \parbox{4mm}{-6} & \parbox{4mm}{6} & \parbox{4mm}{6} & \parbox{4mm}{6} & \parbox{4mm}{6} & \parbox{4mm}{6} & \parbox{4mm}{6} & \parbox{4mm}{6} & \parbox{4mm}{6} & \parbox{4mm}{6} & \parbox{4mm}{6} & \parbox{4mm}{5} & \parbox{4mm}{5}& \parbox{4mm}{5} \\ \cline{2-15}
& \parbox{4mm}{-5} & \parbox{4mm}{6} & \parbox{4mm}{6} & \parbox{4mm}{6} & \parbox{4mm}{6} & \parbox{4mm}{5} & \parbox{4mm}{5} & \parbox{4mm}{5} & \parbox{4mm}{4} & \parbox{4mm}{4} & \parbox{4mm}{4} & \parbox{4mm}{3} & \parbox{4mm}{3} & \parbox{4mm}{3}  \\ \cline{2-15}
& \parbox{4mm}{-4} & \parbox{4mm}{6} & \parbox{4mm}{6} & \parbox{4mm}{6} & \parbox{4mm}{6} & \parbox{4mm}{5} & \parbox{4mm}{5} & \parbox{4mm}{5} & \parbox{4mm}{4} & \parbox{4mm}{3} & \parbox{4mm}{3} & \parbox{4mm}{2} & \parbox{4mm}{2} & \parbox{4mm}{2} \\ \cline{2-15}
& \parbox{4mm}{-3} & \parbox{4mm}{6} & \parbox{4mm}{6} & \parbox{4mm}{6} & \parbox{4mm}{6} & \parbox{4mm}{5} & \parbox{4mm}{5} & \parbox{4mm}{5} & \parbox{4mm}{4} & \parbox{4mm}{3} & \parbox{4mm}{2} & \parbox{4mm}{2} & \parbox{4mm}{1} & \parbox{4mm}{0} \\ \cline{2-15}
& \parbox{4mm}{-2} & \parbox{4mm}{5} & \parbox{4mm}{4} & \parbox{4mm}{4} & \parbox{4mm}{4} & \parbox{4mm}{3} & \parbox{4mm}{3} & \parbox{4mm}{3} & \parbox{4mm}{3} & \parbox{4mm}{2} & \parbox{4mm}{2} & \parbox{4mm}{1} & \parbox{4mm}{0} & \parbox{4mm}{-1} \\ \cline{2-15}
& \parbox{4mm}{-1} & \parbox{4mm}{5} & \parbox{4mm}{4} & \parbox{4mm}{3} & \parbox{4mm}{3} & \parbox{4mm}{2} & \parbox{4mm}{2} & \parbox{4mm}{2} & \parbox{4mm}{2} & \parbox{4mm}{2} & \parbox{4mm}{1} & \parbox{4mm}{0} & \parbox{4mm}{0} & \parbox{4mm}{-1} \\ \cline{2-15}
& \parbox{4mm}{0} & \parbox{4mm}{5} & \parbox{4mm}{4} & \parbox{4mm}{3} & \parbox{4mm}{2} & \parbox{4mm}{2} & \parbox{4mm}{1} & \parbox{4mm}{0} & \parbox{4mm}{0} & \parbox{4mm}{0} & \parbox{4mm}{0} & \parbox{4mm}{-1} & \parbox{4mm}{-1} & \parbox{4mm}{-2} \\ \cline{2-15}
& \parbox{4mm}{1} & \parbox{4mm}{4} & \parbox{4mm}{3} & \parbox{4mm}{2} & \parbox{4mm}{2} & \parbox{4mm}{2} & \parbox{4mm}{1} & \parbox{4mm}{0} & \parbox{4mm}{-1} & \parbox{4mm}{-1} & \parbox{4mm}{-1} & \parbox{4mm}{-2} & \parbox{4mm}{-2} & \parbox{4mm}{-3} \\ \cline{2-15}
& \parbox{4mm}{2} & \parbox{4mm}{3} & \parbox{4mm}{2} & \parbox{4mm}{2} & \parbox{4mm}{1} & \parbox{4mm}{1} & \parbox{4mm}{1} & \parbox{4mm}{0} & \parbox{4mm}{-1} & \parbox{4mm}{-1} & \parbox{4mm}{-1} & \parbox{4mm}{-2} & \parbox{4mm}{-3} & \parbox{4mm}{-4} \\ \cline{2-15}
& \parbox{4mm}{3} & \parbox{4mm}{2} & \parbox{4mm}{2} & \parbox{4mm}{1} & \parbox{4mm}{0} & \parbox{4mm}{0} & \parbox{4mm}{0} & \parbox{4mm}{0} & \parbox{4mm}{-1} & \parbox{4mm}{-1} & \parbox{4mm}{-2} & \parbox{4mm}{-3} & \parbox{4mm}{-4} & \parbox{4mm}{-5} \\ \cline{2-15}
& \parbox{4mm}{4} & \parbox{4mm}{2} & \parbox{4mm}{1} & \parbox{4mm}{0} & \parbox{4mm}{-1} & \parbox{4mm}{-2} & \parbox{4mm}{-2} & \parbox{4mm}{-2} & \parbox{4mm}{-2} & \parbox{4mm}{-2} & \parbox{4mm}{-3} & \parbox{4mm}{-3} & \parbox{4mm}{-4} & \parbox{4mm}{-5} \\ \cline{2-15}
& \parbox{4mm}{5} & \parbox{4mm}{1} & \parbox{4mm}{0} & \parbox{4mm}{0} & \parbox{4mm}{-1} & \parbox{4mm}{-2} & \parbox{4mm}{-3} & \parbox{4mm}{-3} & \parbox{4mm}{-3} & \parbox{4mm}{-3} & \parbox{4mm}{-4} & \parbox{4mm}{-4} & \parbox{4mm}{-4} & \parbox{4mm}{-5} \\ \cline{2-15}
& \parbox{4mm}{6} & \parbox{4mm}{0} & \parbox{4mm}{-1} & \parbox{4mm}{-1} & \parbox{4mm}{-2} & \parbox{4mm}{-3} & \parbox{4mm}{-4} & \parbox{4mm}{-5} & \parbox{4mm}{-5} & \parbox{4mm}{-5} & \parbox{4mm}{-6} & \parbox{4mm}{-6} & \parbox{4mm}{-6} & \parbox{4mm}{-6} \\\hline
\end{tabular}
\end{center}
\end{table}

The computation of the fuzzy feedback scheduling algorithm is relatively simple with
a time complexity of \emph{O}(1), implying that only a small feedback scheduling overhead
has been introduced.
\end{asparaenum}

\section{Performance Evaluation}
\label{sec:5}

To evaluate the performance of the proposed FFS scheme,
this section conducts simulations for the case study system described in Section 4.
The FFS method will be compared with the following two methods:
\begin{itemize}
\item Open-loop scheduling: No feedback scheduling strategy is used; control loops
always run with fixed sampling periods;
\item Ideal feedback scheduling: Actual CPU utilization and all timing
parameters of the tasks are precisely known; and the feedback scheduler
adapts sampling periods using (\ref{equ:3}) and (\ref{equ:4}).
\end{itemize}

The open-loop scheduling method is chosen for comparison because traditionally
it is the default solution for embedded control systems and almost all existing
embedded control systems are based on open-loop scheduling. By comparing FFS with
the ideal feedback scheduling scheme, it is possible to examine how the FFS scheme
will maximize the system performance. The simulation environment is based on
Matlab/TrueTime \cite{R21}. As a performance metric for mobile robot control, the tracking
error is defined as the distance between the robot and the target:
\begin{equation} \label{eq:6}
\begin{split}
ERR_{track}=\sqrt{(x_{act}-x_{ref})^2+(y_{act}-y_{ref})^2}
\end{split}
\end{equation}
where the tuple ($x_{act}$, $y_{act}$) denotes the actual coordinates of the robot, and the tuple ($x_{ref}$, $y_{ref}$) denotes the reference coordinates of the mobile target.

\subsection{Setup overview}
Suppose that the mobile target moves along a half circle with a constant angular
speed. It starts from point (0, 0) at time t = 0, and reaches point (2, 0) at time
t = 4s. The default unit on the plane is meter (m). At runtime,
the actual execution times of three tasks are generated by $c_i=(1+\varepsilon)\bar{c_i}$,
where $\varepsilon$ is a sequence of white Gaussian noise with zero mean and a
variance of 0.01, and $\bar{c_i}$ is the average execution time of each task, as given
in Table 3. The nominal period for each control task is 3, 4 and 5 ms, respectively,
where the period of non-control task 3 (i.e., $h_3$) is fixed. The maximum
allowable sampling periods of two control loops are $h_{max}$ = 7 ms.

\begin{table}[htb!]
\begin{center}
\caption{Average execution times of the tasks.}
\label{table3}
\begin{tabular}{lcccc}
\hline
Time (s) & 0-1 & 1-2 & 2-3 & 3-4 \\
\hline
Control Task 1 (ms) & 0.6 & 1.2 & 1.2 & 1.2 \\
Control Task 2 (ms) & 0.4 & 0.4 & 1.2 & 1.2 \\
Non-control Task 3 (ms) & 1.0 & 2.0 & 2.0 & 1.5 \\
\hline
\end{tabular}
\end{center}
\end{table}

Fixed priorities are utilized in the system. The feedback scheduler is
implemented as a periodic task with the highest priority. The second
highest priority is assigned to the non-control task $\tau_3$, while $\tau_2$
has the lowest priority. In the feedback scheduling loop, the desired CPU
utilization is set to $U_R$ = 85\%. The invocation interval of the feedback
scheduler task is $T_{FS}$ = 20 ms, and its execution time is assumed to be 0.1 ms.
The measured CPU utilization is generated by $\hat{U}=\frac{c_1}{h_1}+\frac{c_2}{h_2}+\frac{c_3}{h_3}+\delta_U$,
where $0\leq \hat{U}\leq1$, and $\delta_U$ is a consequence of zero-mean white
Gaussian noise with a variance of $r^2$.

\subsection{Results and analysis}
Under open-loop scheduling, the tracking trajectory of the mobile
robot and the tracking error are given in Figure 5. In the upper
part of Figure 5, the solid line gives the trajectory of the target
and the centre of circles corresponds to actual position of the
robot. Because the robot is too far from the target, there is no
robot trajectory within the scope of the very right part of the
upper sub-figure.

\begin{figure}[htb!]
\begin{minipage}[t]{0.47\textwidth} \centering
\includegraphics[scale=0.5, bb=98 274 473 573]{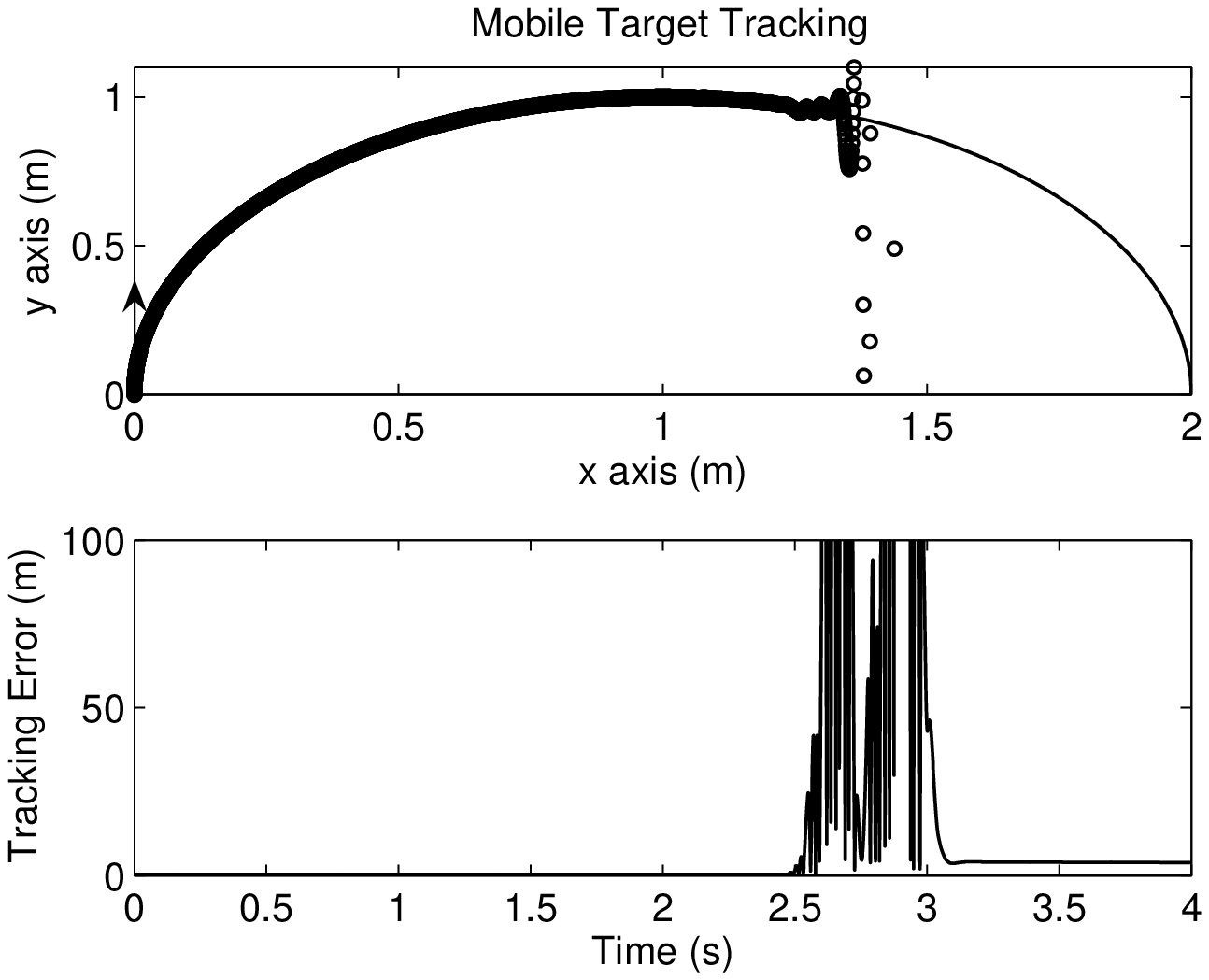}
\caption{Target tracking performance under open-loop scheduling.}
\label{fig:5}
\end{minipage}%
\begin{minipage}[t]{0.47\textwidth}
\includegraphics[scale=0.5, bb=107 275 472 573]{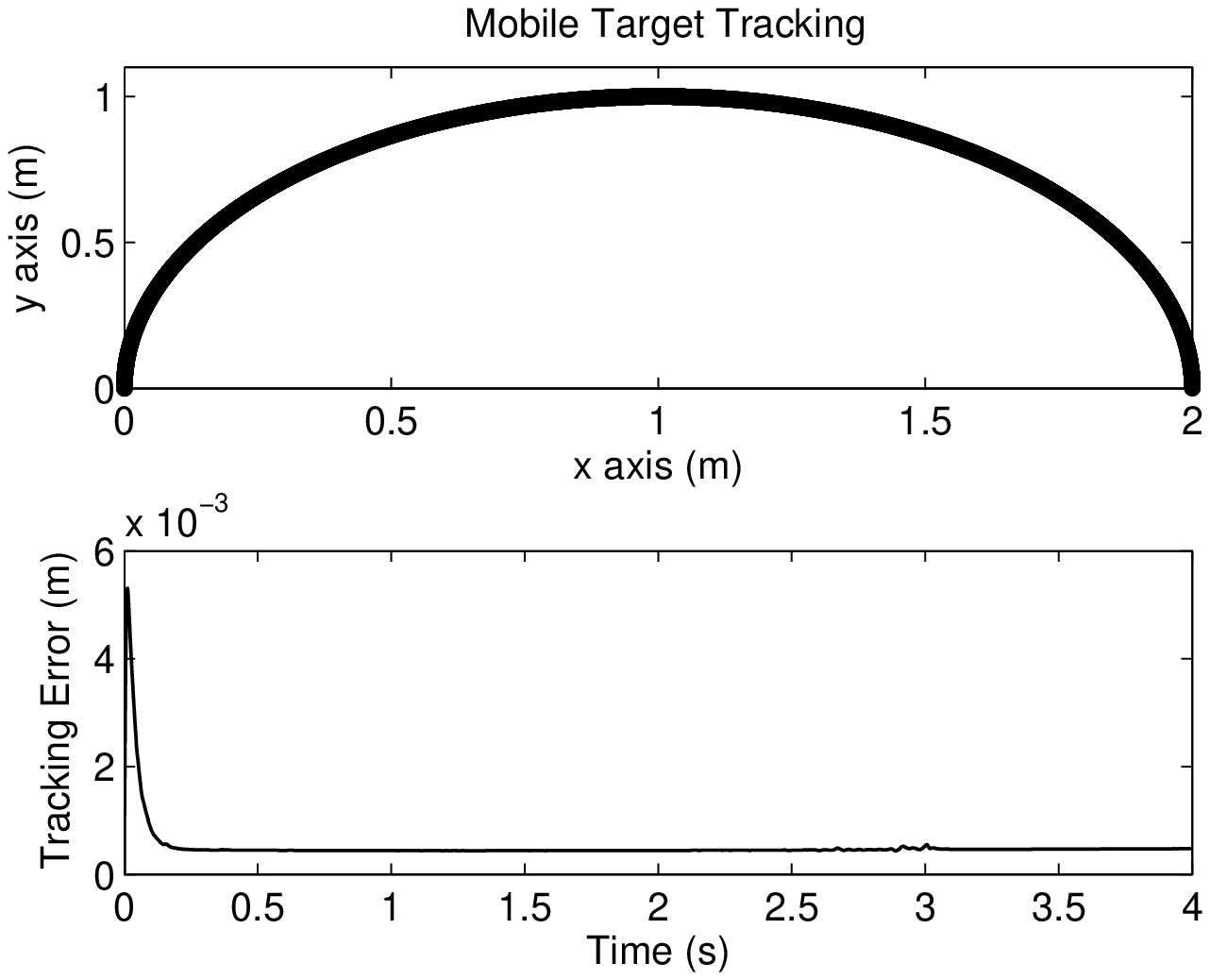}
\caption{Target tracking performance under ideal feedback scheduling.}
\label{fig:6}
\end{minipage}
\end{figure}

After time t = 2s, the robot becomes unable to track the mobile
target effectively, implying that the system finally becomes
unstable. This happens because the system is overloaded from t = 2s
to 3s with the average workload of (1.2/3+1.2/4+2/5)$\times$100\% =
110\%. Since task 2 has the lowest priority, the control loop for
the \emph{y} coordinate of the robot suffers from severe deadline
misses, preventing the robot from well tracking the target.

The performance of the ideal feedback scheduler is shown in Figure
6. It can be seen from Figure 6 that the robot can track the target
very well. The average tracking error throughout the whole
experiment is as small as 0.51mm. A comparison between Figures 6 and
5 reveals that feedback scheduling is effective in dealing with
dynamic variations in system workload and thus enables flexible QoC
management. With feedback scheduling, control performance guarantees
can be achieved for systems operating under changing conditions.

To assess the performance of the FFS scheme, extensive simulations
have been conducted for systems with measurement noises of different
magnitudes. Some representative results are depicted in Figure 7.
The moving tracks of the robot are the same as that in the case of
ideal feedback scheduling and hence are omitted here. Figure 8
depicts the task periods for the case $r$=0.1. Clearly, the fuzzy
feedback scheduler dynamically adjusts sampling periods at runtime,
which is in contrast to the open-loop scheduling scheme that uses
fixed sampling periods.

\begin{figure}[htbp!]
\begin{minipage}[t]{0.47\textwidth} \centering
\includegraphics[scale=0.5, bb=98 274 473 573]{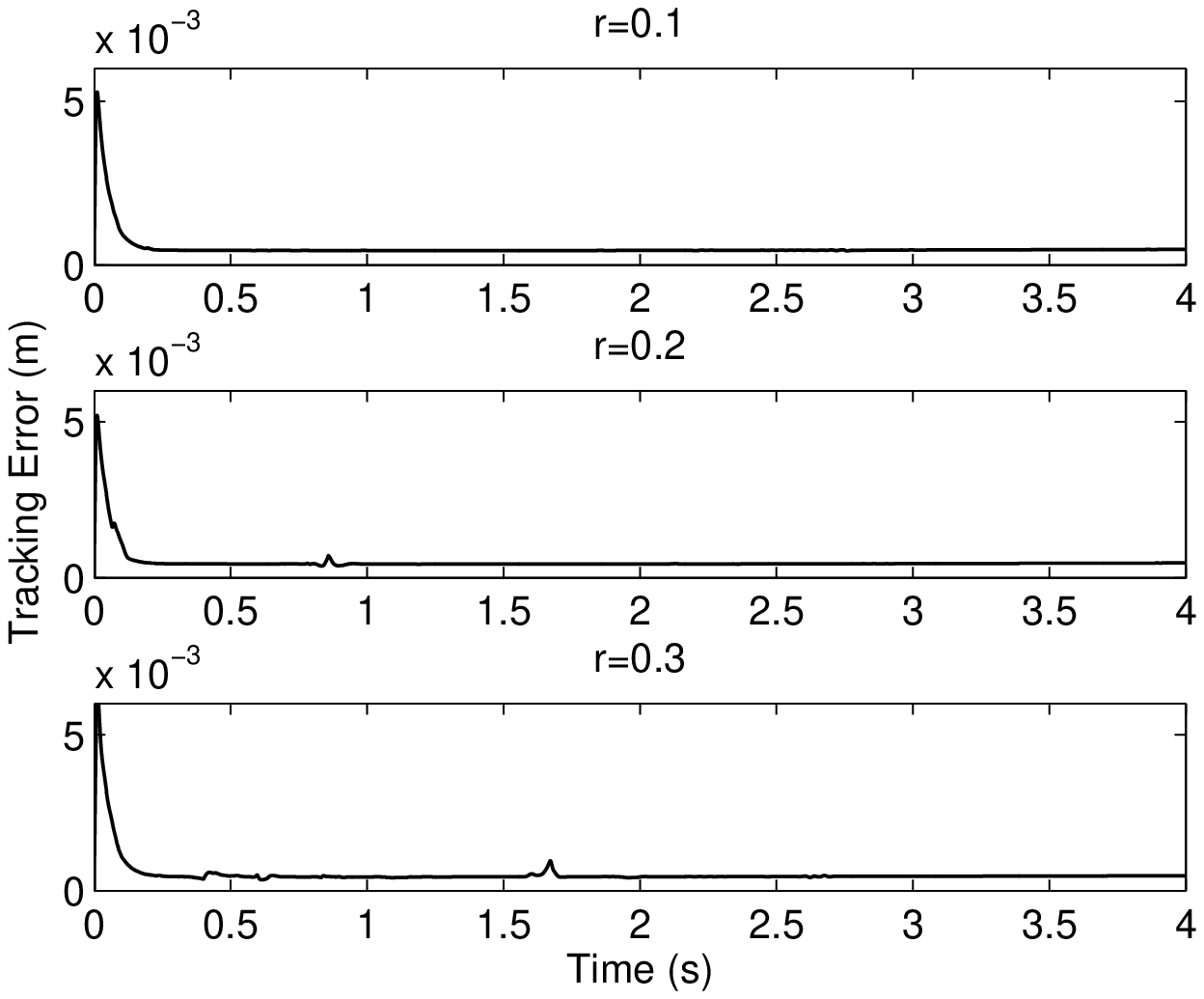}
\caption{Target tracking performance under fuzzy feedback
scheduling} \label{fig:7}
\end{minipage}%
\begin{minipage}[t]{0.47\textwidth}
\includegraphics[scale=0.5, bb=107 275 472 573]{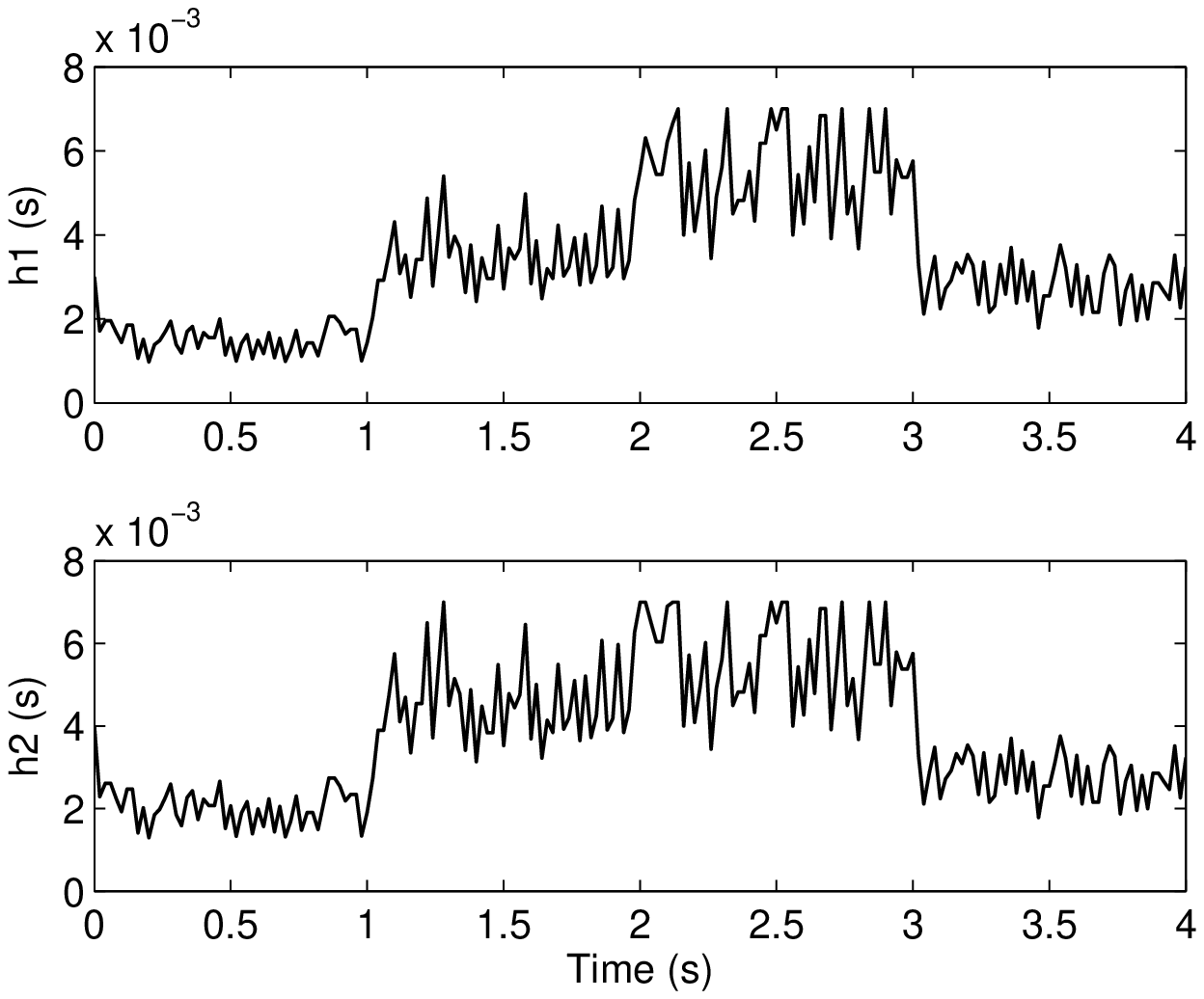}
\caption{Task periods under fuzzy feedback scheduling}
\label{fig:8}
\end{minipage}
\end{figure}

It can be seen from the above results and analysis that: 1) The
proposed FFS scheme is capable of coping with uncertainties in
resource availability; 2) It is robust to measurement noises of
different magnitudes; and 3) It can deliver system performance
comparable to that of the ideal feedback scheduling scheme.

\section{Conclusion}
\label{sec:6}

A fuzzy feedback scheduling scheme has been proposed to deal with
the uncertainty in resource availability in embedded control
systems. Featuring co-design of control and scheduling, the scheme
integrates fuzzy control technology and feedback scheduling. The
framework and the design methodology of the scheme have been
described. Using the look-up table method, an intelligent feedback
scheduling algorithm has been developed which incurs only a small
computational overhead, is easy to implement, and can deal with
measurement noises effectively. Unlike conventional feedback
scheduling methods, the FFS scheme does not rely on the availability
of task execution times and thus is a useful tool in practical
embedded control systems.

\section*{Acknowledgments}
This work is partially supported by the China Postdoctoral Science
Foundation under grant number 20070420232, the Australian Research
Council (ARC) under Discovery Projects grant number DP0559111, the
Australian Government's Department of Education, Science and
Training (DEST) under International Science Linkages grant number
CH070083, and the Natural Science Foundation of China under grant
number 60774060.


\end{document}